\documentstyle[draft,aps]{revtex}

\newtheorem{theorem}{Theorem}
\newtheorem{acknowledgement}[theorem]{Acknowledgement}

\begin{document}
\title{Fast quantum logic gates with trapped ions interacting with external laser
and quantized cavity field beyond the Lamb-Dicke regime}
\author{S. Shelly Sharma$^{1,2}$ and A. Vidiella-Barranco$^{1}$}
\address{(1) Instituto de F\'{i}sica ``Gleb Wataghin'', Universidade Estadual de\\
Campinas, 13083-970 Campinas SP Brazil\\
(2) Depto. de F\'{i}sica, Universidade Estadual de Londrina, 86040-370\\
Londrina, PR Brazil\\
email : shelly@uel.br, vidiella@ifi.unicamp.br}
\maketitle

\begin{abstract}
A scheme to implement quantum logic gates by manipulating trapped ions
through interaction with monochromatic external laser field and quantized
cavity field, beyond the Lamb-Dicke regime, is presented. Characteristic
times, for implementing ionic state transitions using non-resont laser pulse
or quantized cavity field, show a sharp decline for a relatively large
Lamb-Dicke parameter value of $\eta _{L}=\eta _{c}=0.2$ $,$ and are seen to
decrease further with increase in number of initial state vibrational quanta 
$m$.
\end{abstract}

Quantum computation is equivalent to performing a number of unitary operations on a multi qubit quantum system. A single qubit unitary gate and two-qubit controlled-NOT(CN) gate constitute a typical universal set for implementation of multi qubit logic gates \cite{bare95}. Manipulation of trapped cold two-level ions
interacting with laser field \cite{monr95,wine98}, offers a mechanism for realizing unitary operations needed to construct quantum logic gates in Lamb-Dicke regime \cite{cira95} and beyond \cite{wei02,wei202}. Ions in a trap placed inside a high-Q cavity with a single mode quantized field, radiated by an external laser field \cite{Raus99}, offer a much more versatile physical system. The advantage over ion-trap system comes from additional degrees of freedom and the constraints to which ionic motion is subjected to, due to quantization of the cavity field. The state transfer between cavity field mode and motional states of trapped ion has been suggested \cite{park99} and other experiments with ion trap placed inside a cavity have been proposed \cite{semi01,pach02,jane01}. On the experimental side, a trapped ion has been used to probe the cavity field \cite{guth01}, and coherent coupling of an ion to the cavity field has been achieved \cite{mund02}. For experimental realization of quantum logic gates involving several qubits, the ion state manipulation time is an extremely important factor. We propose the
construction of a Hadamard gate, quantum phase gate, and controlled-NOT gate, through manipulation of internal and vibrational states of cold trapped ions placed inside a single mode high-Q cavity. Ionic state manipulation involves selective adjustment
of laser-ion and quantized field-ion detuning parameters. Working in the region beyond the Lamb-Dicke regime ($\eta _{L}<1$ and $\eta _{c}<1$, where $\eta _{L}$ and $\eta_{c}$ are Lamb-Dicke (LD) parameters relative to the external laser and cavity field respectively), we find
considerable reduction in time needed to implement elementary qubit
operations. Implementation of controlled-NOT gate in double Lamb-Dicke regime($\eta _{L}\ll 1$ and $\eta _{c}\ll 1$), has been proposed in ref. \cite{vidi02}.

Consider a two-level ion confined in a trap placed inside a high-finesse
cavity. The ion is radiated by the single mode cavity field of frequency $%
\omega _{c}$ and an external laser field of frequency $\omega _{L}$ and
phase $\phi $. Entanglement of internal states of the ion, vibrational states of ionic center of mass, and the states of quantized cavity field, through interaction of ion with external laser
and the cavity field, provides basic mechanism for qubit
state rotations in Hilbert space. The system Hamiltonian, for the case when center of the trap is close to the node of cavity field standing wave, is given by \cite{vidi02} 
\begin{equation}
\hat{H}=\hat{H}_{0}+\hat{H}_{int},  \label{eq1}
\end{equation}
\begin{equation}
\hat{H}_{0}=\hbar \nu \left( \hat{a}^{\dagger }\hat{a}+\frac{1}{2}\right)
+\hbar \omega _{c}\hat{b}^{\dagger }\hat{b}+\frac{\hbar \omega _{0}}{2}%
\sigma _{z},  \label{eq2}
\end{equation}
\begin{eqnarray}
\hat{H}_{int} &=&\hbar \Omega \left[ \sigma _{+}\exp \left[ i\eta _{L}(\hat{a%
}^{\dagger }+\hat{a})-i(\omega _{L}t+\phi )\right] +h.c.\right]  \nonumber \\
&&+\hbar g(\sigma _{+}+\sigma _{-})\left( \hat{b}^{\dagger }+\hat{b}\right)
\sin \left[ \eta _{c}(\hat{a}^{\dagger }+\hat{a})\right] ,  \label{eq3}
\end{eqnarray}
where $\hat{a}^{\dagger }(\hat{a})$ and $\hat{b}^{\dagger }(\hat{b})$ are
creation(destruction) operators for vibrational phonon and cavity field
photon respectively and $\omega _{0}$ the transition frequency of the
two-level ion. The ion-phonon and ion-cavity coupling constants are $\Omega $
and $g$, whereas $\sigma _{k}(k=z,+,-)$ are the Pauli operators qualifying
the internal state of the ion. 

In the interaction picture, determined by unitary transformation $%
U_{0}(t)=\exp \left[ -{i\hat{H_{0}}t}/{\hbar }\right] $, the Hamiltonian $%
\hat{H_{I}}$ reads as 
\begin{eqnarray}
\hat{H}_{I} &=&\hbar \Omega \lbrack \sigma _{+}{\hat{O}_{0}^{L}}\exp \left[
i\left( \delta _{0L}t-\phi \right) \right] +h.c.]  \nonumber \\
&+&\hbar \Omega \left[ \sigma _{+}\sum_{k=1}^{\infty }(i\eta _{L})^{k}{%
\hat{O}_{k}^{L}\hat{a}}^{k}\exp \left[ i\left( (\delta _{0L}-k\nu )t-\phi
\right) \right] +h.c.\right]   \nonumber \\
&+&\hbar \Omega \left[ \sigma _{+}\sum_{k=1}^{\infty }(i\eta _{L})^{k}{%
\hat{a}^{\dagger k}}{\hat{O}_{k}^{L}}\exp \left[ i\left( (\delta _{0L}+k\nu
)t-\phi \right) \right] +h.c.\right]   \nonumber \\
&+&\hbar g\left[ \sigma _{+}\hat{b}^{\dagger }\sum_{k=1,3,..}^{\infty
}(i^{k-1}{\eta _{c}}^{k}){\hat{a}^{\dagger k}}{\hat{O}_{k}^{c}}\exp \left[
i(\delta _{0c}+k\nu +2\omega _{c})t\right] +h.c.\right]   \nonumber \\
&+&\hbar g\left[ \sigma _{+}\hat{b}^{\dagger }\sum_{k=1,3,..}^{\infty
}\left( i^{k-1}{\eta _{c}}^{k}\right) {\hat{O}_{k}^{c}}\hat{a}^{k}\exp \left[
i(\delta _{0c}-k\nu +2\omega _{c})t\right] +h.c.\right]   \nonumber \\
&+&\hbar g\left[ \sigma _{+}\hat{b}\sum_{k=1,3,..}^{\infty }\left( i^{k-1}{%
\eta _{c}}^{k}\right) {\hat{O}_{k}^{c}}\hat{a}^{k}\exp \left[ i(\delta
_{0c}-k\nu )t\right] +h.c.\right]   \nonumber \\
&+&\hbar g\left[ \sigma _{+}\hat{b}\sum_{k=1,3,..}^{\infty }\left( i^{k-1}{%
\eta _{c}}^{k}\right) {\hat{a}^{\dagger k}}{\hat{O}_{k}^{c}}\exp \left[
i(\delta _{0c}+k\nu )t\right] +h.c.\right] ,  \label{eq4}
\end{eqnarray}
where $\delta _{0L}=\omega _{0}-\omega _{L}$, $\delta _{0c}=\omega
_{0}-\omega _{c}$, and 
\begin{equation}
{{\hat{O}}_{k}}=\exp \left( -\frac{\eta ^{2}}{2}\right) \sum_{p=0}^{\infty }%
\frac{(i\eta )^{2p}\hat{a}^{\dagger p}\hat{a}^{p}}{p!\left( p+k\right) !}%
\text{ .}  \label{eqop}
\end{equation}
The matrix element of diagonal operator ${{\hat{O}}_{k}}$ for a given
vibrational state $m$ is given by 
\begin{equation}
\left\langle m\left| {\hat{O}_{k}}\right| m\right\rangle =\exp (-\frac{\eta
^{2}}{2})\sum\limits_{p=0}^{m}\frac{(i\eta _{L})^{2p}m!}{p!\left( p+k\right)
!(m-p)!}.  \label{eqopm}
\end{equation}
We may note that the case where center of the trap lies close to the
anti node of cavity standing wave field can be treated in an analogous
fashion. 

We next examine the construction of quantum logic gates namely Hadamard
gate, phase gate, and Controlled-NOT gate by proper choice of detuning
parameters $\delta _{0L}$ and $\delta _{0c}$, and phase $\phi $ for ionic
state manipulation. For this purpose, we consider the time evolution of the
system for the special choices (i) $\delta _{0L}=0$ that is the resonant
laser field, (ii) $\delta _{0L}=k\nu $, non resonant laser pulse, and (iii) $%
\delta _{0c}=-k\nu $ where the quantized cavity field is resonant with $%
k^{th}$ blue-shifted vibrational side band.

{\Large Case I. Resonant laser Pulse, }$\delta _{0L}=0$

In rotating wave approximation, the relevant part of Hamiltonian is 
\begin{equation}
\hat{H_{1}}=\hbar \Omega \lbrack \sigma _{+}{\hat{O}_{0}^{L}\exp (-i\phi
_{1})}+\sigma _{-}{\hat{O}_{0}^{L}\exp (i\phi _{1})}].  \label{eq5}
\end{equation}
We work in the basis $\left| g,m,n\right\rangle ,\left| e,m,n\right\rangle ,$
where $m,n=0,1,..,\infty $, denote the state of ionic vibrational motion and
quantized cavity field, respectively. The state of the system at a time $t,$
starting from initial states $\left| g,m,n\right\rangle $, and $\left|
e,m,n\right\rangle $, is given respectively by 
\begin{equation}
\left| g,m,n\right\rangle \Rightarrow \cos \left( \Omega F_{m,m}^{L}t\right)
\left| g,m,n\right\rangle -i{\exp (-i\phi _{1})}\sin \left( \Omega
F_{m,m}^{L}t\right) \left| e,m,n\right\rangle ,  \label{eq6}
\end{equation}
and 
\begin{equation}
\left| e,m,n\right\rangle \Longrightarrow \cos \left( \Omega
F_{m,m}^{L}t\right) \left| e,m,n\right\rangle -i{\exp (i\phi _{1})}\sin
\left( \Omega F_{m,m}^{L}t\right) \left| g,m,n\right\rangle ,  \label{eq7}
\end{equation}
where the real matrix element $F_{m,m}^{L}=\left\langle m\left| {\hat{O}%
_{0}^{L}}\right| m\right\rangle $. In the approximation, $\eta _{L}\ll 1,$ $%
F_{m,m}^{L}$ approaches $1$ for all values of $m,$ whereas for $\eta _{L}<1$
in the region beyond the Lamb-Dicke regime, the matrix element $%
F_{m,m}^{L}<1 $.

{\Large Case II. Non resonant laser pulse, }$\delta _{0L}=k\nu ${\Large , }$%
k=1,2,..\infty $

The time evolution, in rotating wave approximation, is governed by the term 
\begin{equation}
\hat{H_{2}}=\hbar \Omega \left[ \sigma _{+}\left( {i\eta _{L}}\right) ^{k}{%
\hat{O}_{k}^{L}}{\hat{a}^{k}}{\exp (-i\phi _{2})}+\sigma _{-}\left( {-i\eta
_{L}}\right) ^{k}{\hat{a}^{\dagger k}}{\hat{O}_{k}^{L}}{\exp (i\phi _{2})}%
\right] .  \label{eq8}
\end{equation}
The initial states $\left| g,m,n\right\rangle $, and $\left|
e,m,n\right\rangle $, evolve as 
\begin{eqnarray}
\left| g,m,n\right\rangle &\Rightarrow &\cos \left[ \Omega F_{m-k,m}^{L}t%
\right] \left| g,m,n\right\rangle  \nonumber \\
&&-i{\exp (-i\phi _{2})}\sin \left[ \Omega F_{m-k,m}^{L}t\right] \left|
e,m-k,n\right\rangle ,\text{ for}\hspace{0.2cm}m\geq k;  \label{eq9a} \\
\left| g,m,n\right\rangle &\Rightarrow &\left| g,m,n\right\rangle ,\text{
for}\hspace{0.2cm}m<k,  \label{eq9} \\
\left| e,m,n\right\rangle &\Rightarrow &\cos \left[ \Omega F_{m+k,m}^{L}t%
\right] \left| e,m,n\right\rangle  \nonumber \\
&&-i{\exp (i\phi _{2})}\sin \left[ \Omega F_{m+k,m}^{L}t\right] \left|
g,m+k,n\right\rangle \text{.} \label{eq10}
\end{eqnarray}
Here the matrix element $F_{m-k,m}^{L}$ stands for 
\begin{equation}
F_{m-k,m}^{L}=i^{k}\left[ \prod_{i=0}^{k-1}\eta _{c}\sqrt{(m-i)}\right]
\left\langle m-k\left| {\hat{O}_{k}^{L}}\right| m-k\right\rangle ,
\label{eq11}
\end{equation}
and $F_{m,m-k}^{L}=\left( F_{m-k,m}^{L}\right) ^{\ast }.$

{\Large Case III. Quantized field in resonance with }
$ k^{th}${\Large \ blue
shifted vibrational sideband, }$ \delta _{0c}=-k\nu $
{\Large , }$k=1,3,..\infty $.

Neglecting quickly vibrating terms the Hamiltonian reduces to 
\begin{equation}
\hat{H_{3}}=\hbar g\left[ \sigma _{+}\hat{b}\left( i^{k-1}{\eta _{c}^{k}}%
\right) {\hat{a}^{\dagger k}}{\hat{O}_{k}^{c}}+\sigma _{-}\hat{b}^{\dagger
}\left( \left( -i\right) ^{k-1}{\eta _{c}^{k}}\right) {\hat{O}_{k}^{c}\hat{a}%
^{k}}\right] \text{.}  \label{eq12}
\end{equation}
From the initial states $\left| g,m,n\right\rangle $, and $\left|
e,m,n\right\rangle $, the system evolves as 
\begin{eqnarray}
\left| g,m,n\right\rangle &\Rightarrow &\cos \left[ g\sqrt{n}F_{m+k,m}^{c}t%
\right] \left| g,m,n\right\rangle  \nonumber \\
&&-i\sin \left[ g\sqrt{n}F_{m+k,m}^{c}t\right] \left| e,m+k,n-1\right\rangle
,  \label{eq13}
\end{eqnarray}
\begin{eqnarray}
\left| e,m,n\right\rangle &\Rightarrow &\cos \left[ g\sqrt{n+1}F_{m-k,m}^{c}t%
\right] \left| e,m,n\right\rangle  \nonumber \\
&&-i\sin \left[ g\sqrt{n+1}F_{m-k,m}^{c}t\right] \left|
g,m-k,n+1\right\rangle \text{ for}\hspace{0.2cm}m\geq k;  \label{eq14} \\
\left| e,m,n\right\rangle &\Rightarrow &\left| e,m,n\right\rangle ,\text{
for}\hspace{0.2cm}m<k,  \label{eq15}
\end{eqnarray}
with the matrix element, $F_{m-k,m}^{c}=\langle m-k|i^{k-1}{\eta _{c}^{k}%
\hat{O}_{k}^{c}}\hat{a}^{k}|m\rangle $ given by 
\begin{equation}
F_{m-k,m}^{c}=(i^{k-1})\left[ \prod_{i=0}^{k-1}\eta _{c}\sqrt{(m-i)}\right]
\left\langle m-k|\hat{O}_{k}^{c}|m-k\right\rangle .  \label{eq16}
\end{equation}
For the special case $k=1$, 
\begin{equation}
F_{m-1,m}^{c}=\exp \left( -\frac{\eta _{c}^{2}}{2}\right) \eta _{c}\sqrt{m}%
\sum_{p=0}^{m-1}\frac{(i\eta _{c})^{2p}\left( m-1\right) !}{\left(
p+1\right) !p!\left( m-1-p\right) !}.  \label{eq17}
\end{equation}

We notice that for a given value of $\eta _{c}$, the characteristic
frequencies, $\left[ g\sqrt{\left( n+1\right) }F_{m-k,m}^{c}\right] $ and $%
\left[ g\sqrt{n}F_{m+k,m}^{c}t\right] $ vary with the number of initial
state vibrational and quantized field quanta. Choosing $m,$ $n$ and $\eta
_{c}$ to give large value for the matrix element $F_{m-k,m}^{c}$, can 
reduce the time required for implementing a quantum logic operation
involving states $\left| e,m,n\right\rangle $ and $\left|
g,m-k,n+1\right\rangle $. We may recall here that with the center of trap
close to standing wave anti node, transitions involve state change by $k$
quanta, where $k=0,2,4,...,\infty $.

{\Large Implementation of quantum logic gates}

A Hadamard gate rotates single qubit states as below 
\begin{equation}
|0\rangle \Longrightarrow \frac{|0\rangle +|1\rangle }{\sqrt{2}},
\hspace{0.2in}%
|1\rangle \Longrightarrow \frac{|0\rangle -|1\rangle }{\sqrt{2}}.
\label{eq18}
\end{equation}
A controlled-NOT gate on the other hand transforms two-qubit states as 
\begin{eqnarray}
|0,0\rangle &\Longrightarrow &|0,0\rangle \text{ ,}\hspace{0.2in}|0,1\rangle
\Longrightarrow |0,1\rangle ,  \nonumber \\
|1,0\rangle &\Longrightarrow &|1,1\rangle \text{ ,\hspace{0.2in}}|1,1\rangle
\Longrightarrow |1,0\rangle ,  \label{eq19}
\end{eqnarray}
i.e., target qubit changes (does not change) state if the control qubit
(vibrational state) has value one (zero). An application of a Hadamard gate,
followed by a phase gate and another application of a Hadamard gate results
in the implementation of a controlled-NOT gate. The phase gate involving a
special phase shift of $\pi $, results in the following elementary
operation, 
\begin{eqnarray}
|0,0\rangle &\Longrightarrow &|0,0\rangle \text{ ,}\hspace{0.2in}|0,1\rangle
\Longrightarrow |0,1\rangle ,  \nonumber \\
|1,0\rangle &\Longrightarrow &|1,0\rangle \text{ ,\hspace{0.2in}}|1,1\rangle
\Longrightarrow -|1,1\rangle .  \label{eq20}
\end{eqnarray}

With the cavity initially prepared in the vacuum state consider the initial
states $\left| g,m,0\right\rangle $ and $\left| e,m,0\right\rangle $. The
internal state of the ion is the target qubit whereas state of center of
mass motion acts as control qubit. Choosing $\delta _{0L}=0$, $\phi _{1}=%
\frac{\pi }{2}+2\pi p,(p=1,2..)$ and applying a laser pulse for time $%
t_{m,m} $, such that $\left[ \Omega F_{m,m}^{L}t_{m,m}\right] =\frac{\pi }{4}%
,$ we get from Eqs. (\ref{eq6} and \ref{eq7}) 
\begin{equation}
\left| g,m,0\right\rangle \Rightarrow \frac{1}{\sqrt{2}}\left( \left|
g,m,0\right\rangle -\left| e,m,0\right\rangle \right) ,  \label{eq21}
\end{equation}
\begin{equation}
\left| e,m,0\right\rangle \Rightarrow \frac{1}{\sqrt{2}}\left( \left|
e,m,0\right\rangle +\left| g,m,0\right\rangle \right) .  \label{eq22}
\end{equation}
This is the first step to implement a Hadamard gate. Next, a non-resonant pulse with $%
\delta _{0L}=k\nu ,k=1,2...$, 
interacting with the ion for time, $t_{m+k,m}={\pi }/(\Omega F_{m+k,m}^{L})$, may be used to switch the sign of state $\left| e,m,0\right\rangle $ for $m<k$, as verified from Eqs. (\ref{eq9a}- \ref{eq10}) and shown by Wei et al \cite{wei202}. Tuning the quantized cavity field to $k^{th}$ blue shifted vibrational sideband by choosing $\delta _{0c}=-k\nu$,$%
k=1,3,..\infty $, for an interaction time, 
$ t_{m-k,m}={%
\pi }/(gF_{m-k,m}^{c}) $, switches the sign of state $\left| e,m,0\right\rangle$  with $m \geq k$ (Eqs. (\ref{eq13} and \ref{eq14}). The end result of this two step process is implementation of a Hadamard gate, 

\begin{equation}
\left| g,m,0\right\rangle \Rightarrow \frac{1}{\sqrt{2}}\left( \left|
g,m,0\right\rangle +\left| e,m,0\right\rangle \right) ,  \label{eq23}
\end{equation}
\begin{equation}
\left| e,m,0\right\rangle \Rightarrow \frac{1}{\sqrt{2}}\left( \left|
g,m,0\right\rangle -\left| e,m,0\right\rangle \right) .  \label{eq24}
\end{equation}
As the dependence of the matrix elements 
$F_{m,m}^{L}$, $F_{m+k,m}^{L}$ and $F_{m-k,m}^{c}$  on Lamb-Dicke parameters varies with $m$ and $k$, these choices
determine the implementation time for Hadamard gate, which is $t_{H}=t_{m,m}+ t_{m+k,m}$ for $m<k$, and $t_{H}=t_{m,m}+ t_{m-k,m}$ for $m\geq k$.

To implement the phase gate, choose $\delta _{0c}=-k\nu $, and the
interaction time of the ion with the cavity field to be $ t_{P}={%
\pi }/(gF_{m-k,m}^{c}) $. From Eqs. (\ref{eq13}-\ref{eq15}), we find
that 
\begin{eqnarray}
\left| g,m,0\right\rangle &\Longrightarrow &\left| g,m,0\right\rangle ;%
\hspace{0.2in}\text{for\hspace{0.2cm}all }\hspace{0.2cm}m,  \nonumber \\
\left| e,m,0\right\rangle &\Longrightarrow &\left| e,m,0\right\rangle ;%
\hspace{0.2in}\text{for}\hspace{0.2cm}m<k,  \nonumber \\
\left| e,m,0\right\rangle &\Longrightarrow &-\left| e,m,0\right\rangle ;%
\hspace{0.2in}\text{for}\hspace{0.2cm}m\geq k,  \label{eq25}
\end{eqnarray}
or the phase gate of Eq. (\ref{eq20}) is implemented. The quantized cavity field states serve as the auxiliary qubit in the implementation of phase gate required to implement a controlled-NOT quantum logic gate. An important observation is that the the cavity is initially in the vacuum state and is left in the vacuum state in the end \cite{vidi02}. As such successive quantum logic gates can be implemented even in the presence of cavity losses.

The speed of operation, in ionic state manipulation, is extremely important
for successful experimental realization. The operation time should not be so
long as to allow the decoherence due to interaction with environment to
dominate the scene. To analyze the switching times, we consider the
control-bit vibrational phonons to take values $m$ and $m-1$ ($m=1,4,9,$ and 
$16).$ Table I and II list the expected ion-laser/quantized field
interaction times $t_{m,m}$ and $t_{m-1,m}$ for $\Omega =g=2\pi \times 140$
kHz and the choices $\eta _{L}=\eta _{c}=0.02$ (LD regime), and $0.2$
(beyond LD regime).\ Experimentally cold ion traps with parameter values
given above have been realized successfully \cite{monr95} and ions in
vibrational state with $m$ up to $16$ obtained \cite{wine98}. In experiments
performed with trapped Calcium ions and optical cavities \cite{guth01},
ion-cavity coupling $g\approx 6MHz.$

In the Lamb-Dicke regime ($\eta _{L}=\eta _{c}=0.02$) , the characteristic
time, $t_{m,m}^{1}$, needed during the realization of a Hadamard gate is
independent of $m$ since $F_{m,m}\approx 1.$ Working beyond the Lamb-Dicke
limit ($\eta _{L}=\eta _{c}=0.2$ here), $F_{m,m}<1,$ causes an increase in $%
t_{m,m}^{2}$ with increasing $m$ . The last column in table I shows that for 
$m=16,$ $t_{16,16}^{2}$ is twice as large as $t_{1,1}^{1}.$ We may consider $%
2\mu s$ to be the characteristic time for the step involving interaction of
ion with resonant laser pulse.

The characteristic time, $ t_{m-1,m}={\pi }/(gF_{m-1,m}) $, for
implementing ionic state transition using off resonance laser pulse or
quantized cavity field, is seen to decrease with increase in the number of
initial state quanta $m,$ for $\eta _{L}=\eta _{c}=0.02$, as well as $0.2.$
In addition for the case $\eta _{L}=\eta _{c}=0.2$ $,$ that is going out
from LD regime, $t_{m-1,m}$ shows a sharp decline for all $m$ values
considered. In general for a given $m$ value $t_{m-1,m}$ is larger in
comparison with $t_{m,m}.$ \ Usually the control qubit is encoded by
vibrational quanta $0$ and $1$. We notice however that working beyond the LD
regime, large $m$ values result in shorter times for ionic state
manipulation. For example the choice $m=16,$ results in ${t_{0,1}^{1}}/{%
t_{15,16}^{2}}=28.5$. It is exciting to think that by choosing hotter
initial ionic states, the state manipulation time can come down from $%
\approx 178\mu s$ to $\approx 6\mu s$. We conclude that faster ionic state manipulation for implementing quantum logic gates is achieved in ion trap - optical cavity system, by working beyond
the Lamb-Dicke regime rather than in Lamb-Dicke regime. Recently a coherence time of $\approx 1$ ms for qubits based on single Ca$^+$ ion has been reported \cite{schm03}. We must recall however that the
optical cavity decay time is, $\tau _{c}\approx 2\mu s$ \cite{mund02,walt00}. In comparison, the ionic state manipulation
times for operations involving change of cavity state,  are still quite large and much needs to be done in the experimental field.   \bigskip\ 

\begin{acknowledgement}
S.S.S would like to thank Unicamp for hospitality. This work is partially
supported by CNPq, Brazil.
\end{acknowledgement}

\newpage 
\begin{table}[tbp] \centering%
%
\caption{Characteristic times, $\left( t_{m,m}={\pi }/(4\Omega F_{m,m})\right) ,$ 
and matrix elements $F_{m,m}$ for implementing ionic state transitions using resonant laser pulse/ quantized 
cavity field, for Rabi frequency, $\Omega =g=2\pi \times 140 kHz$.\label{t1}}
\begin{tabular}[b]{cccccc}
\multicolumn{3}{c}{LD regime, $\eta _{L}=\eta _{c}=0.02$} & 
\multicolumn{3}{c}{Beyond LD regime, $\eta _{L}=\eta _{c}=0.2$} \\ 
$m$ & $F_{m,m}$ & $t_{m,m}^{1}\left( \mu s\right) $ & $F_{m,m}$ & $%
t_{m,m}^{2}\left( \mu s\right) $ & $R_{m,m}$=$\frac{t_{1,1}^{1}}{t_{m,m}^{2}}
$ \\ 
$1$ & $0.999$ & $0.90$ & $0.941$ & $0.95$ & $0.9$ \\ 
$4$ & $0.998$ & $0.90$ & $0.828$ & $1.07$ & $0.8$ \\ 
$9$ & $0.996$ & $0.90$ & $0.654$ & $1.36$ & $0.7$ \\ 
$16$ & $0.993$ & $0.90$ & $0.441$ & $2.02$ & $0.5$%
\end{tabular}
\end{table}%
%

\bigskip

\begin{table}[tbp] \centering%
%
\caption{Characteristic times, $\left( t_{m-1,m}={\pi }/(gF_{m-1,m})\right) $, 
and matrix elements $F_{m-1,m}$ for implementing ionic state transitions 
using off resonance laser pulse/ quantized 
cavity field for Rabi frequency, $\Omega =g=2\pi \times 140 kHz$.\label{t2}} 
\begin{tabular}{cccccc}
\multicolumn{3}{c}{LD regime, $\ \eta _{L}=\eta _{c}=0.02$} & 
\multicolumn{3}{c}{Beyond LD regime, $\ \eta _{L}=\eta _{c}=0.2$} \\ 
$m$ & $F_{m-1,m}$ & $t_{m-1,m}^{1}\left( \mu s\right) $ & $F_{m-1,m}$ & $%
t_{m-1,m}^{2}\left( \mu s\right) $ & $R_{m-1,m}$=$\frac{t_{0,1}^{1}}{%
t_{m-1,m}^{2}}$ \\ 
$1$ & $0.02$ & $178.57$ & $0.196$ & $18.22$ & $9.8$ \\ 
$4$ & $0.04$ & $89.28$ & $0.369$ & $9.68$ & $18.4$ \\ 
$9$ & $0.06$ & $59.52$ & $0.498$ & $7.17$ & $24.9$ \\ 
$16$ & $0.08$ & $44.64$ & $0.570$ & $6.26$ & $28.5$%
\end{tabular}
\end{table}%
%

\end{document}